\def\hybrid{\topmargin 0pt      \oddsidemargin 0pt
        \headheight 0pt \headsep 0pt
        \textwidth 160true mm       
        \textheight 231true mm         
        \marginparwidth 0.0in
        \parskip 0pt plus 1pt   \jot = 1.5ex}
\catcode`\@=11
\def\marginnote#1{}

\newcount\hour
\newcount\minute
\newtoks\amorpm
\hour=\time\divide\hour by60
\minute=\time{\multiply\hour by60 \global\advance\minute by-\hour}
\edef\standardtime{{\ifnum\hour<12 \global\amorpm={am}%
        \else\global\amorpm={pm}\advance\hour by-12 \fi
        \ifnum\hour=0 \hour=12 \fi
        \number\hour:\ifnum\minute<10 0\fi\number\minute\the\amorpm}}
\edef\militarytime{\number\hour:\ifnum\minute<10 0\fi\number\minute}

\def\draftlabel#1{{\@bsphack\if@filesw {\let\thepage\relax
   \xdef\@gtempa{\write\@auxout{\string
      \newlabel{#1}{{\@currentlabel}{\thepage}}}}}\@gtempa
   \if@nobreak \ifvmode\nobreak\fi\fi\fi\@esphack}
        \gdef\@eqnlabel{#1}}
\def\@eqnlabel{}
\def\@vacuum{}
\def\draftmarginnote#1{\marginpar{\raggedright\scriptsize\tt#1}}

\def\draft{\oddsidemargin -.5truein
        \def\@oddfoot{\sl preliminary draft \hfil
        \rm\thepage\hfil\sl\today\quad\militarytime}
        \let\@evenfoot\@oddfoot \overfullrule 3pt
        \let\label=\draftlabel
        \let\marginnote=\draftmarginnote
   \def\@eqnnum{(\theequation)\rlap{\kern\marginparsep\tt\@eqnlabel}%
\global\let\@eqnlabel\@vacuum}  }

\relax
\hybrid

\documentclass[12pt]{article}
\def\be{\begin{equation}}\def\ee{\end{equation}}\def\ba{\begin{eqnarray}}\def\ea{\end{eqnarray}}
\def\lb{\label}
\newcommand{\bea}{\begin{eqnarray}}\newcommand{\eea}{\end{eqnarray}}
\newcommand{\p}[1]{(\ref{#1})}
\newcommand{\tb}{{\tilde b}}\newcommand{\tc}{{\tilde c}}

\def\theequation{\thesection.\arabic{equation}}

\begin{document}

\renewcommand{\thefootnote}{\fnsymbol{footnote}}

\thispagestyle{empty}
\vspace*{3cm}
\begin{center}{\Large\bf BRST operators for $W$ algebras}\end{center}
\vspace{1cm}

\begin{center}
{\large\bf A.P.~Isaev${}^{a,}$\footnote{Corresponding author}, S.O.~Krivonos${}^{a}$ and
O.V.~Ogievetsky${}^{b}$ }
\end{center}

\begin{center}
${}^{a}$ Bogoliubov Laboratory of Theoretical Physics, \\
Joint Institute for Nuclear Research, \\
Dubna, Moscow region 141980, Russia \\
E-mail: isaevap@theor.jinr.ru, krivonos@theor.jinr.ru
\\
\vspace{0.3cm}
${}^{b}$ Center of Theoretical Physics\footnote{Unit\'e Mixte de Recherche
(UMR 6207) du CNRS et des Universit\'es Aix--Marseille I,
Aix--Marseille II et du Sud Toulon -- Var; laboratoire affili\'e \`a la FRUMAM (FR 2291)}, Luminy,
13288 Marseille, France \\
and P. N. Lebedev Physical Institute, Theoretical Department,
Leninsky pr. 53, 117924 Moscow, Russia \\
E-mail: oleg@cpt.univ-mrs.fr
\end{center}
\vspace{2cm}

\begin{abstract}
The study of quantum Lie algebras motivates a use of non-canonical ghosts and anti-ghosts
for non-linear algebras, like $W$-algebras.
This leads, for the $W_3^{\phantom{(2)}}\!\! $ and $W_3^{(2)}$ algebras, to the BRST operator
having the conventional cubic form.
\end{abstract}

\newpage
\setcounter{page}{1}
\renewcommand{\thefootnote}{\arabic{footnote}}
\setcounter{footnote}{0}
\section{Introduction}
The BRST symmetry was discovered in \cite{brs} and \cite{t} in the context of
gauge theories. The symmetry is generated by the BRST operator (BRST charge) $Q$ whose square
is zero, $Q^2=0$. It is used to describe the physical state space in constrained Hamiltonian
(and Lagrangian) systems (see \cite{GT}, \cite{HT} and references therein).
The BRST operator $Q$ is related to the homology of Lie algebras (\cite{bcr}, \cite{vh}).

\vskip .3cm

The simplest BRST operator appears in theories with first class constraints $\chi_i$,
$i=1,2,\dots$; the constraints satisfy the Lie (super-)algebra relations
\be\lb{suli}
\chi_{i_1}\, \chi_{i_2}-(-1)^{(i_1)(i_2)}\, \chi_{i_2}\, \chi_{i_1} =
C^{k_1}_{i_1 i_2} \, \chi_{k_1}\ .\ee
Here $(k)\equiv 0,1\ {\mathrm{mod}}\, 2\ $ is the parity of the generator $\chi_k$, $(k)\equiv 0$
for a boson constraint $\chi_k$ and  $(k)\equiv 1$ for a fermion constraint $\chi_k$;
$C^{k_1}_{i_1 i_2}$ are the structure constants which satisfy the Jacobi identity together with
\be\lb{strc1} C^k_{mn}=0\ \ \ {\mathrm{if}}\ (k)+(n)+(m)\equiv\!\!\!\!\! /\ 0\; \ee
and
\be\lb{strc2}  C^k_{mn}=(-)^{(m)(n)+1}\, C^k_{nm}\; .\ee
To prepare the stage for general {\it quantum Lie algebras} (see the definition in Section 2) we rewrite (\ref{suli}) in the form
\be\lb{suli1}
\chi_{i_1}\, \chi_{i_2}-\sigma^{k_1 k_2}_{i_1 i_2}\, \chi_{k_1}\, \chi_{k_2} =
C^{k_1}_{i_1 i_2} \, \chi_{k_1} \ ,\ee
where
\be\lb{superp}\sigma^{k_1 k_2}_{i_1 i_2}=(-1)^{(i_1)(i_2)}\delta^{k_1}_{i_2}\delta^{k_2}_{i_1}\;
\ee
is the super-permutation matrix. For the quantum Lie algebras, the operator $\sigma$ is an
invertible solution of the Yang--Baxter equation which is not necessarily the super-per\-mu\-tation.
We shall see in Section 2 how to write the Jacobi identity and the conditions (\ref{strc1}) and
(\ref{strc2}) in the general case.

\vskip .3cm
The BRST operator for the algebra (\ref{suli}) is cubic in generators; it has the standard
"two-term" form
\be\lb{suliQ}
Q=c^n\chi_{n}-\frac{1}{2}\, c^n\, c^m\,\phi_{mn}^{rs}\, C_{rs}^k b_k \; ,
\ee
where the additional generators $\{ c^n, \; b_m \}$ are the canonical (super) ghosts and
anti-ghosts; we shall write the relations concerning the ghost---anti-ghost sector in the form
adapted for the general quantum Lie algebras. The relations between ghosts---anti-ghosts and
the generators $\chi_k$ are
\be\lb{cangh1}b_{m}\,\chi_{n}=\phi_{mn}^{kl}\, \chi_{k}\, b_{l}\; ,\;\;\;
\chi_{m}\, c^{n}=c^{l}\,\phi_{lm}^{kn}\, \chi_{k}\; ,\ee
where
\be\lb{suliph1}\phi_{mn}^{kl} = (-1)^{(n)((m)+1)} \delta^k_n \delta^l_m \; .\ee
The relations for ghosts---anti-ghosts are
\be\lb{cangh2}\begin{array}{c}
b_{i_1}\, b_{i_2}=-\tilde{\sigma}_{i_1 i_2}^{j_1 j_2}\, b_{j_1}\, b_{j_2}\; ,\;\;\;
c^{k_2}\, c^{k_1}=-c^{j_2}\, c^{j_1}\tilde{\sigma}_{j_1 j_2}^{k_1 k_2}\; , \\ [0.3cm]
b_{i_2}\, c^{k_2}=-c^{j_1}\, (\tilde{\sigma}^{-1})_{j_1 i_2}^{n_1 k_2}\, b_{n_1}+
\delta_{i_2}^{k_2}\; ,\\ [0.3cm]
\end{array}\ee
where
\be\lb{suliph2}\tilde{\sigma}_{mn}^{kl}=(-1)^{(m)(n)+(m)+(n)}\delta^k_n\delta^l_m\ .\ee
Note that
\be\lb{suliph3}\tilde{\sigma}_{mn}^{kl}=\phi_{mn}^{pq}\sigma^{rs}_{pq}(\phi^{-1})_{rs}^{kl}
\qquad{\mathrm{or}}\qquad\tilde{\sigma}=\phi\,\sigma\,\phi^{-1}\ .
\ee
In (\ref{cangh1}) and (\ref{cangh2}) we have taken fermionic (respectively, bosonic)
ghosts---anti-ghosts $c^k$ and $b_k$ for the bosonic (respectively, fermionic) constraints
$\chi_k$. There is however an alternative choice\footnote{With this choice, the BRST operator has
the same form (\ref{suliQ}). The relation between two choices is discussed in \cite{IsOgGo},
Section 6.} of the ghost sector in the algebra: one can set the matrices $\phi$ and $\tilde{\sigma}$
to be the super-permutations (\ref{superp}), $$\phi=\tilde{\sigma}=\sigma\ .$$
The relation (\ref{suliph3}) trivially holds for this choice as well.

\vskip .3cm
The fact of having several choices for
(\ref{cangh1}), (\ref{cangh2}) is a strong motivation for us to work with not necessarily
canonical ghosts in the construction of the BRST operators for general quadratic algebras.

\vskip .3cm
In this paper we discuss the form of the BRST operator for the
gauge theories with constraints satisfying the quadratic relations (\ref{suli1}) with the braid
matrix $\sigma$ which is not necessarily equal to the (super-)permutation matrix. We restrict our
attention to the simplest (unitary) situation when $\sigma^2=1$. The BRST operator was
investigated in the framework of 2D conformal theories for various non-linear algebras which
include the Virasoro algebra (like $W$-type algebras), see, e.g, \cite{TM}, \cite{ssn},
\cite{ikeda}. The interest in the construction of the BRST charge $Q$ for quadratic algebras
(\ref{suli1}) has been recently renewed in \cite{BuchLav, BuchLav2, BuchLav3} in the study of the 
higher spins in the AdS spaces (about quadratic algebras in this context see \cite{MVas}). The 
main idea of our approach is to modify the ghost sector in the algebra in a way, compatible with 
the algebra of constraints (and not insisting on having the canonical ghosts, as it is done 
usually, see, e.g., \cite{ssn} and \cite{BuchLav}). More precisely, for the algebra defined by the
quadratic relations (\ref{suli1}) with a given braid matrix $\sigma$, we work with the quadratic
ghost sector whose defining relations may contain the matrix $\sigma$ as well.

\vskip .3cm
The paper is organized as follows. In Section 2 we introduce non-canonical quadratic ghosts and
anti-ghosts and generalize the construction of the BRST operator for Lie (super-)algebras
to quantum Lie algebras with the unitary braid matrix $\sigma$.
The data needed to define the ghost sector include a {\it twist pair} $\{\sigma,\phi\}$ of
braid matrices. In Section 3 we consider two physical examples of
quadratic algebras -- the $W_3$ and $W_3^{(2)}$ algebras.
As in section 2, we work with general
quadratic ghost algebras. For the $W_3$ algebra, it turns out that the most general ghost sector
includes two arbitrary parameters.
We show that the BRST operator exists for arbitrary values of the parameters.
The known BRST operators for the $W_3$ and $W_3^{(2)}$ algebras, based on the usual, canonical,
ghosts and anti-ghosts, contain terms of degree higher than 3 in the generators.
We find that for a certain choice of the values of the parameters (entering the definition of the
ghost sector), the BRST operator has the simple conventional cubic form as in the case of the Lie
(super-)algebras (\ref{suliQ}).

\setcounter{equation}0
\section{BRST operator for quadratic algebras}

Consider an algebra ${\cal{A}}$ with generators $\chi_i$ and defining quadratic--linear relations
of the form
\be\chi_{i_1}\,\chi_{i_2}-\sigma^{k_1k_2}_{i_1i_2}\,\chi_{k_1}\,\chi_{k_2}=
C^{k_1}_{i_1 i_2}\,\chi_{k_1}\ee
or, in the concise notation (\cite{FRT}, \cite{IsOg2}),
\be\lb{qlaK}\chi_{1\rangle}\,\chi_{2\rangle}-\sigma_{12}\,\chi_{1\rangle}\,\chi_{2\rangle}=
C^{\langle 1}_{12\rangle}\,\chi_{1\rangle}\; .\ee
The algebra ${\cal{A}}$ is called {\it quantum Lie algebra} (QLA) if
the structure constants $\sigma^{ij}_{kl}$ and $C^k_{ij}$ satisfy (for details see \cite{Ber}, 
\cite{IsOg2} and references therein)
\be\lb{int1b}\sigma^{j_1j_2}_{i_1i_2}\,\sigma^{p_2k_3}_{j_2i_3}\,\sigma^{k_1k_2}_{j_1p_2}=
\sigma^{j_2j_3}_{i_2i_3}\,\sigma^{k_1p_2}_{i_1j_2}\,\sigma^{k_2 k_3}_{p_2 j_3}\; ,\ee
\be\lb{int2b}C^{p_1}_{n_1n_2}\, C^{k_4}_{p_1n_3}=\sigma^{p_2p_3}_{n_2n_3} \,
C^{k_1}_{n_1p_2}\, C^{k_4}_{k_1p_3}+C^{p_3}_{n_2n_3}\, C^{k_4}_{n_1p_3}\; ,\ee
\be\lb{int3b}C^{p_1}_{n_1n_2}\,\sigma^{k_1k_3}_{p_1n_3}=\sigma^{p_2p_3}_{n_2n_3}\,
\sigma^{k_1j_2}_{n_1p_2}\, C^{k_3}_{j_2p_3}\; ,\ee
\be\lb{int3c}(\sigma^{j_2p_3}_{n_2n_3}\, C^{p_1}_{n_1j_2}+\delta^{p_1}_{n_1} C^{p_3}_{n_2n_3})\,
\sigma^{k_1k_3}_{p_1p_3}=\sigma^{p_1p_2}_{n_1n_2}\, (\sigma^{j_2k_3}_{p_2n_3}\,
C^{k_1}_{p_1j_2}+\delta^{k_1}_{p_1}\, C^{k_3}_{p_2n_3})\; ,\ee
\be\lb{int4c}\exists\ \ t^i_{jk}:\ C^i_{jk}=(\delta^l_j\delta^m_k-\sigma^{lm}_{jk})t^i_{lm}\ .\ee

\noindent
Eq. (\ref{int1b}) is the braid relation for the matrix $\sigma^{kl}_{ij}$ and (\ref{int2b}) is
an analogue of the Jacobi identity. Eq. (\ref{int3b}) reduces to
(\ref{strc1}) for Lie super-algebras. In the concise notation (\cite{FRT}, \cite{IsOg2}), the
identities (\ref{int1b}) -- (\ref{int4c}) read
\be\lb{int1a}\sigma_{12}\,\sigma_{23}\,\sigma_{12}=\sigma_{23}\,\sigma_{12}\,\sigma_{23}\; ,\ee
\be\lb{int2}C^{\langle 1}_{12\rangle }\, C^{\langle 4}_{13\rangle }=\sigma_{23}\,
C^{\langle 1}_{12\rangle }\, C^{\langle 4}_{13\rangle }+C^{\langle 3}_{23\rangle }\,
C^{\langle 4}_{13\rangle }\; ,\ee
\be\lb{int3}C^{\langle 1}_{12\rangle }\,\sigma_{13}=\sigma_{23}\,\sigma_{12}\,
C^{\langle 3}_{23\rangle }\; ,\ee
\be\lb{int3a}(\sigma_{23}\, C^{\langle 1}_{12\rangle }+C^{\langle 3}_{23\rangle })\,
\sigma_{13}=\sigma_{12}\, (\sigma_{23}\, C^{\langle 1}_{12\rangle }+
C^{\langle 3}_{23\rangle } ) \; ,\ee
\be\lb{int4}C^{\langle 1}_{12\rangle } = (1-\sigma_{12}) t^{\langle 1}_{12\rangle } \; .\ee

Below we consider the simplest, unitary, braid matrices $\sigma$, that is,
\be\lb{s2eq1}\sigma_{nm}^{pj}\sigma^{ki}_{pj}=\delta^k_n\delta^i_m\ \ {\mathrm{or}}\ \
\sigma^2 = 1\; .\ee
In this situation, eq. (\ref{int3a}) follows from eq. (\ref{int3}) and eq. (\ref{int4})
is equivalent to (cf. (\ref{strc2}) for Lie super-algebra case)
\be\lb{add2}(1+\sigma_{12})C^{\langle 1}_{12\rangle}=0\; .\ee

Introduce the ghost generators $c^i$ and $b_i$ with defining quadratic relations
\be\lb{qlabc}b_{1\rangle}\, b_{2\rangle}=-\tilde{\sigma}_{12}\, b_{1\rangle}\, b_{2\rangle}
\; , \;\;\; c^{\langle 2}\, c^{\langle 1}=-c^{\langle 2}\, c^{\langle 1}\tilde{\sigma}_{12}\; ,\ee
\be\lb{qlabcK}b_{2\rangle }\, c^{\langle 2}=-c^{\langle 1}\,\tilde{\sigma}^{-1}_{12}\,
b_{1\rangle }+I_2\; ,\ee
where $I$ is the identity operator. We require that the matrix $\tilde{\sigma}^{ij}_{kl}$
satisfies the braid relation as well,
\be\label{int5a}\tilde{\sigma}_{12}\,\tilde{\sigma}_{23}\,\tilde{\sigma}_{12}=
\tilde{\sigma}_{23}\,\tilde{\sigma}_{12}\,\tilde{\sigma}_{23}\; ,\ee
which ensures that different reorderings of monomials $b_{1\rangle }b_{2\rangle }c^{\langle 2}$,
$b_{2\rangle }c^{\langle 2}c^{\langle 1}\;$ {\it etc.} give the same result. We note (and we shall
not repeat it later) that the braid relation is stronger than the reordering requirement .

\vskip .3cm
The appearance of another braid matrix $\tilde{\sigma}$ in (\ref{qlabc}), (\ref{qlabcK}) follows
the example of Lie super-algebras (see eqs. (\ref{cangh1}) and (\ref{cangh2})) where the braid
matrix $\sigma$ in (\ref{qlaK}) is the super-permutation matrix (\ref{superp}) while
$\tilde{\sigma}$ is defined in (\ref{suliph2}). Note that in particular one could identify the
braid matrices $\tilde{\sigma}$ and $\sigma$. The possibility $\tilde{\sigma}=\sigma$ is also
motivated by the differential calculi on quantum groups \cite{Wor}, \cite{IsOg2}. Non-linear
algebras with $\tilde{\sigma}=\sigma$ (for general non-unitary matrices $\sigma$) and the
corresponding BRST operators were studied in \cite{IsOg2}, \cite{IsOgGo} and \cite{IsOg3}.

\vskip .3cm
The smash product $\Omega$ of the QLA (\ref{qlaK}) and the ghost algebra (\ref{qlabc}),
(\ref{qlabcK}) is fixed by the cross-commutation relations
\be\lb{crossKK}b_{1\rangle }\,\chi_{2\rangle}=\phi_{12}\,\chi_{1\rangle}\, b_{2\rangle }\; ,
\;\;\;\chi_{2\rangle }\, c^{\langle 2}=c^{\langle 1}\,\phi_{12}\,\chi_{1\rangle }\; .\ee
We require that the matrix $\phi^{ij}_{kl}$ satisfies relations
\be\lb{cons1}\tilde{\sigma}_{12}\,\phi_{23}\,\phi_{12}=\phi_{23}\,\phi_{12}\, \tilde{\sigma}_{23}
\; ,\;\;\phi_{12}\,\phi_{23}\,\sigma_{12}=\sigma_{23}\,\phi_{12}\,\phi_{23}\; ,\;\;\ee
\be\lb{cons3}\phi_{12}\phi_{23}C_{12\rangle}^{\langle 1}\delta_{3\rangle}^{\langle 2}=
C_{23\rangle}^{\langle 2}\,\phi_{12}\; ,\ee
which ensure that different reorderings of monomials $b_{1\rangle }\chi_{2\rangle }c^{\langle 2}$,
$b_{1\rangle }b_{2\rangle}\chi_{3\rangle}$, $b_{1\rangle}\chi_{2\rangle}\chi_{3\rangle }\,$
{\it etc.} give the same result.

\vskip .3cm
We now construct the BRST operator for the algebra $\Omega$.

\vspace{0.3cm}\noindent {\bf Proposition.} {\it Let
\be Q_{gh}:=-\frac{1}{2}\, c^{\langle 2}\, c^{\langle 1}\;
\phi_{12}C^{\langle 1}_{12\rangle}\, b_{1\rangle }\ .\ee
The operator (cf. (\ref{suliQ}))
\be\lb{brstK}Q=c^{\langle 1}\chi_{1\rangle}+Q_{gh}\in\Omega\ee
satisfies
\be\lb{zero}Q^2=0\ee
if the matrix $\phi$ obeys (\ref{cons3}) and defines the twist
between the matrices $\sigma$ and $\tilde{\sigma}$:
\bea\lb{twist}&\tilde{\sigma}_{12}=\phi_{12}\,\sigma_{12}\,\phi_{12}^{-1}\; ,&\\[.3em]
\lb{cons2}&\sigma_{12}\,\phi_{23}\,\phi_{12}=\phi_{23}\,\phi_{12}\,\sigma_{23}\; ,\;\;
\phi_{12}\,\phi_{23}\,\sigma_{12}=\sigma_{23}\,\phi_{12}\,\phi_{23}\; ,&\\[.4em]
\lb{cons2'}&\phi_{12}\,\phi_{23}\,\phi_{12}=\phi_{23}\,\phi_{12}\,\phi_{23}\; .&\eea
}

\vspace{0.3cm}\noindent {\bf Proof.} We note that (\ref{int5a}) and (\ref{cons1}) follow from (or
are contained in) (\ref{int1a}), (\ref{twist}) and (\ref{cons2}). In view of (\ref{s2eq1}) and
(\ref{twist}) we have $\tilde{\sigma}^2=1$.

\paragraph{1.} First, $Q_{gh}^2=0$. Indeed,
\be\label{caa}\begin{array}{c}4\, Q_{gh}^2=c^{\langle 4}\, c^{\langle 3}\phi_{34}
C^{\langle 3}_{34 \rangle}b_{3\rangle }c^{\langle 3}\, c^{\langle 2}\phi_{23}
C^{\langle 2}_{23\rangle}b_{2\rangle }\\[1em]
=c^{\langle 4}\, c^{\langle 3}\phi_{34}C^{\langle 3}_{34\rangle}\left( c^{\langle 2}
\, c^{\langle 1}\,\tilde{\sigma}_{23}^{-1}\tilde{\sigma}_{12}^{-1}b_{1\rangle}+c^{\langle 2}
(1-\tilde{\sigma}_{23}^{-1})\right)\phi_{23}C^{\langle 2}_{23\rangle}b_{2\rangle }\\[1em]
=c^{\langle 4}\cdots c^{\langle 1}\phi_{34}C^{\langle 3}_{34\rangle}
\tilde{\sigma}_{23}\tilde{\sigma}_{12}\phi_{23}C^{\langle 2}_{23\rangle}b_{1\rangle}b_{2\rangle }
+c^{\langle 3}\, c^{\langle 2}c^{\langle 1}\phi_{23}C^{\langle 2}_{23\rangle}
\phi_{12}(1-\sigma_{12})C^{\langle 1}_{12\rangle}b_{1\rangle }\\[1em]
=c^{\langle 4}\cdots c^{\langle 1}\phi_{1234}\sigma_{23}\sigma_{12}\sigma_{34}
\sigma_{23}C^{\langle 3}_{34\rangle}\phi_{23}^{-1}\phi_{12}^{-1}C^{\langle 2}_{23\rangle}
b_{1\rangle}b_{2\rangle }+2c^{\langle 3}\, c^{\langle 2}c^{\langle 1}\phi_{123}
C^{\langle 1}_{12\rangle}\delta_{3\rangle}^{\langle 2}C^{\langle 1}_{12\rangle}b_{1 \rangle}\\[1em]
=c^{\langle 4}\cdots c^{\langle 1}\phi_{1234}C^{\langle 3}_{34\rangle}
C^{\langle 1}_{12\rangle}\delta_{3\rangle}^{\langle 2}\phi_{12}^{-1}b_{1\rangle}b_{2\rangle}
+2c^{\langle 3}\, c^{\langle 2}c^{\langle 1}\phi_{123}C^{\langle 1}_{12\rangle}
\delta_{3\rangle}^{\langle 2}C^{\langle 1}_{12\rangle}b_{1 \rangle}=0\; .\end{array}\ee
In the second equality we used (\ref{qlabcK}); in the third equality we used $\tilde{\sigma}^2=1$
and relabeled the spaces in the second term; in the fourth equality we introduced, for $i<j$, the
higher-rank matrices
$$\phi_{i,\dots,j} = (\phi_{i,i+1} \phi_{i+1,i+2} \dots \phi_{j-1,j})
(\phi_{i,i+1} \dots \phi_{j-2,j-1}) \cdots (\phi_{i,i+1} \phi_{i+1,i+2})\phi_{i,i+1} \; , $$
which satisfy, by (\ref{twist})--(\ref{cons2'}),
\be\phi_{i,\dots,j} \sigma_{i+k} = \tilde{\sigma}_{j-k-1} \phi_{i,\dots,j} \qquad
{\mathrm{for}}\quad k=0,1, \dots ,j-i-1\; .\lb{ches}\ee
Here $\sigma_l$ stands for $\sigma_{l,l+1}$ (same for $\tilde{\sigma}$).
Then we took into account that, by (\ref{twist})--(\ref{cons2'}), (\ref{int3}) and (\ref{cons3}),
$$\begin{array}{c}\phi_{34}C^{\langle 3}_{34\rangle}\tilde{\sigma}_{23}\tilde{\sigma}_{12}
\phi_{23}=\phi_{34}C^{\langle 3}_{34\rangle}\phi_{23}\sigma_{23}\phi_{23}^{-1}\tilde{\sigma}_{12}
\phi_{23}=\phi_{234}C^{\langle 2}_{23\rangle}\delta_{4\rangle}^{\langle 3}\sigma_{23}
\phi_{23}^{-1}\tilde{\sigma}_{12}\phi_{23}\\[1em]
=\phi_{234}\sigma_{34}\sigma_{23}C^{\langle 3}_{34\rangle}\phi_{23}^{-1}\tilde{\sigma}_{12}
\phi_{23}=\phi_{234}\sigma_{34}\sigma_{23}C^{\langle 3}_{34\rangle}
\phi_{12}\tilde{\sigma}_{23}\phi_{12}^{-1}\\[1em]
=\phi_{234}\sigma_{34}\sigma_{23}\phi_{12}\phi_{23}\phi_{34}C^{\langle 2}_{23\rangle}
\delta_{4\rangle}^{\langle 3}\sigma_{23}\phi_{23}^{-1}\phi_{12}^{-1}=\phi_{1234}\sigma_{23}
\sigma_{12}\sigma_{34}\sigma_{23}C^{\langle 3}_{34\rangle}\phi_{23}^{-1}\phi_{12}^{-1}\;
\end{array}$$
for the first term; we used (\ref{cons3}) and (\ref{add2}) for the second term. The fifth
equality in (\ref{caa}) uses (\ref{cons3}) as well as (\ref{ches}) and
\ba &c^{\langle k}\cdots c^{\langle 1}=c^{\langle k}\cdots c^{\langle 1}\, A_k^{(\tilde{\sigma})}
\; ,&\lb{ches2}\\[.3em]  &A_k^{(\tilde{\sigma})}\,\tilde{\sigma}_j=-A_k^{(\tilde{\sigma})}\quad
\forall\ j<k\; ,&\lb{ches3}\ea
where $A_k^{(\sigma)}$ are the anti-symmetrizing projectors, $(A_k^{(\sigma)})^2=A_k^{(\sigma)}$,
for the braid matrix $\sigma$ constructed with the help of the recurrence relation
$$A_{k+1}^{(\sigma)}=\frac{1}{k!}(1-\sigma_k+\sigma_{k-1}\sigma_k-\dots +(-1)^k\sigma_{1}\dots
\sigma_k)A_k^{(\sigma)}\; ,\;\;\; A_{1}^{(\sigma)} = 1 \; .$$
In the last, sixth, equality in (\ref{caa}) we used (\ref{ches2}) as well as
$$b_{1\rangle}\cdots b_{k\rangle}=A_k^{(\tilde{\sigma})}b_{1\rangle}\cdots b_{k\rangle}\ ,$$
then (\ref{ches}) and then finally the identities (see \cite{IsOg4} for details)
\be\begin{array}{c}\lb{impo}A_4^{(\sigma)}C^{\langle 3}_{34\rangle}C^{\langle 1}_{12\rangle}
\delta_{3\rangle}^{\langle 2}(1-\sigma_1)=0\; ,\\[1em]
A_3^{(\sigma)}C^{\langle 1}_{12\rangle}
\delta_{3\rangle}^{\langle 2}C^{\langle 1}_{12\rangle}=0\; .\end{array}\ee

\paragraph{2.} Since $Q_{gh}^2 =0$, the verification of $Q^2=0$ reduces to
\be\lb{q0}Q^2\equiv (c^{\langle 2}\chi_{2\rangle})^2+[ c^{\langle 1}\chi_{1\rangle},Q_{gh}]_{_+}
\stackrel{?}{=} 0\; ,\ee
where $[ ,]_{_+}$ is the anti-commutator and "?" denotes a statement to be verified.
So we need to check only terms containing the generators $\chi_k$:
\be\lb{q1}c^{\langle 2}\chi_{2\rangle}c^{\langle 2}\chi_{2\rangle}-\frac{1}{2}\left( c^{\langle 1}
\chi_{1 \rangle}c^{\langle 2}c^{\langle 1}\phi_{12}C^{\langle 1}_{12\rangle}b_{1\rangle}+
c^{\langle 2}c^{\langle 1}\phi_{12}C^{\langle 1}_{12\rangle}b_{1\rangle}c^{\langle 1}
\chi_{1\rangle}\right) \stackrel{?}{=}0\; .\ee
{}For the first term in l.h.s. of (\ref{q1}) we have
\be\lb{q2}c^{\langle 2}\chi_{2\rangle}c^{\langle 2}\chi_{2\rangle}=c^{\langle 2}c^{\langle 1}
\phi_{12}\chi_{1\rangle}\chi_{2\rangle}=\frac{1}{2}c^{\langle 2}c^{\langle 1}(1-\tilde{\sigma}_1)
\phi_{12}\chi_{1\rangle}\chi_{2\rangle}=\frac{1}{2}c^{\langle 2}c^{\langle 1}\phi_{12}
C^{\langle 1}_{12\rangle}\chi_{1 \rangle}\ ,\ee
where we have used (\ref{crossKK}), (\ref{qlabc}), (\ref{twist}) and (\ref{qlaK}).
{}For the second term in l.h.s. of  (\ref{q1}) we have
\be\lb{q3}\begin{array}{c}-\frac{1}{2}\left( c^{\langle 3}\chi_{3\rangle}c^{\langle 3}c^{\langle 2}
\phi_{23}C^{\langle 2}_{23\rangle}b_{2\rangle}+c^{\langle 3}c^{\langle 2}\phi_{23}
C^{\langle 2}_{23\rangle}b_{2\rangle}c^{\langle 2}\chi_{2\rangle}\right)\\ [0.3cm]
=-\frac{1}{2}\left( c^{\langle 3}c^{\langle 2}c^{\langle 1}\phi_{23}\phi_{12}\phi_{23}
C^{\langle 2}_{23 \rangle}\phi_{12}^{-1}b_{1\rangle}\chi_{2\rangle}-\frac{1}{2}c^{\langle 3}
c^{\langle 2}\phi_{23} C^{\langle 2}_{23\rangle}(-c^{\langle 1}\tilde{\sigma}_{12}^{-1}
b_{1\rangle}+I_2)\chi_{2 \rangle}\right)\\ [0.3cm]
=-\frac{1}{2}\, c^{\langle 3}c^{\langle 2}c^{\langle 1}\left(\phi_{123}C^{\langle 2}_{23 \rangle}
\phi_{12}^{-1}-\phi_{23}C^{\langle 2}_{23\rangle}\tilde{\sigma}_{12}\right) b_{1\rangle}
\chi_{2\rangle}-\frac{1}{2}\, c^{\langle 3}c^{\langle 2}\phi_{23}C^{\langle 2}_{23\rangle}\,
\chi_{2\rangle}\\ [0.3cm]
=-\frac{1}{2}\, c^{\langle 3}c^{\langle 2}c^{\langle 1}\phi_{123}\left( C^{\langle 2}_{23 \rangle}
-\sigma_{23}\sigma_{12}C^{\langle 2}_{23\rangle}\right)\phi_{12}^{-1}b_{1\rangle}\chi_{2\rangle}
-\frac{1}{2}\, c^{\langle 3}c^{\langle 2}\phi_{23}C^{\langle 2}_{23\rangle}\,\chi_{2\rangle} \; .
\end{array}\ee
In the first equality we used (\ref{crossKK}) and (\ref{qlabc}); in the second equality we used
$\tilde{\sigma}^2=1$; in the third equality we used (\ref{cons3}), (\ref{cons2'}) and \ref{int3}).

\vskip .1cm
Substituting (\ref{q2}) and (\ref{q3}) into the expression for $Q^2$, we obtain
$$Q^2=-\frac{1}{2}\, c^{\langle 3}c^{\langle 2}c^{\langle 1}\phi_{123}\left( 1-\sigma_{23} \sigma_{12}\right) C^{\langle 2}_{23\rangle}\phi_{12}^{-1}b_{1\rangle}\chi_{2\rangle} =0\; .$$
Here we used (\ref{ches2}) and took into account that
\be\lb{q4}
A_3^{(\tilde{\sigma})}\phi_{123}\left( 1-\sigma_{23}\sigma_{12}\right)
=A_3^{(\tilde{\sigma})} \left( 1-\tilde{\sigma}_{12}\tilde{\sigma}_{23}\right)\phi_{123}
=0\; \ee
by (\ref{ches}) and (\ref{ches3}).\hfill $\bullet$

\vspace{0.3cm}
We conclude that the data for the construction of the algebra with ghosts include the twist
pair $\{\sigma ,\phi\}$, like in the study of the quantum matrix algebras and characteristic
equations for them in \cite{IOP}.

\vspace{0.3cm}
As for the quantum matrix algebras, there are two natural possibilities to choose the twisting
matrix $\phi$ for the general braid matrix $\sigma$. The first possibility is
$\phi =\sigma^{\pm 1}$. The second possibility is to choose $\phi$ to be the
super-permutation matrix (\ref{superp}).
Then the ghosts simply (anti-)commute with the generators of the QLA (with this choice of $\phi$,
the twist equations (\ref{cons2}) say only that the braid matrix $\sigma$ should be even, i.e.,
$\sigma^{ij}_{kl} =0$ if $(i)+(j)+(k)+(l) \neq 0$, see, e.g., \cite{IsOb}).
{}For the purely even (bosonic) algebra ${\cal{A}}$, this second choice is the permutation matrix,
$\phi^{kl}_{nm}=\delta^k_m \delta^l_n$, which leads to the direct tensor product of the QLA and
the ghost algebra (\ref{qlabcK}); the ghosts commute with the generators of the QLA,
\be\label{crossK}b_i\,\chi_j=\chi_j\,b_i\; ,\;\;\; c^i\,\chi_j=\chi_j\, c^i\; .\ee

\vskip .2cm
The possibility (\ref{crossK}) will be employed in the next Section for the algebras $W_3$ and
$W_3^{(2)}$ (which have an infinite number of generators).

\vskip .2cm
The bosonic ghost number operator $G$ is well defined in the full algebra of constraints, ghosts 
and anti--ghosts, 
\be [G,\chi_j ]=0\ ,\ [G,b_j]=-b_j\ ,\ [G,c^j]=c^j\qquad {\mathrm{for\ all}}\ \ j\ee
($[x,y]:=xy-yx$ is the usual commutator; the BRST operator $Q$ satisfies $[G,Q]=Q$), so one can 
induce representations from vacuum vectors and build Fock spaces in the usual manner.

\vskip .2cm
{}For a given braid matrix $\sigma^{ij}_{kl}$ (with $\sigma^2 =1$), structure constants $C^i_{jk}$ 
and a twisting matrix $\phi^{ij}_{kl}$, the BRST operator $Q$ of ghost number 1 (i.e. $[G,Q]=Q$) 
is defined uniquely by the requirement of having the conventional form 
$Q=c^i\chi_i+\ {\mathrm{ghost\ terms}}$.

\setcounter{equation}0
\section{BRST operator for $W_3$ and $W_3^{(2)}$ algebras}

We have seen in the previous Section that the BRST charge for quadratic algebras
may be constructed with not necessarily canonical ghosts and anti-ghosts.
The natural question is whether,
in the known examples, the structure of the BRST charge simplifies
for a certain modification of the ghost---anti-ghost sector.

\vskip .3cm
In this Section we consider two simplest quadratic algebras with an infinite number
of generators,
the $W_3$ and $W_3^{(2)}$ algebras. Their known BRST charges, based on the usual, canonical,
ghosts and anti-ghosts, include terms of degree higher than 3 in the generators. We shall see that
there is a freedom in the ghost---anti-ghost sector for these algebras. We demonstrate that for a
certain choice of the ghost---anti-ghost sector, the BRST charges for these algebras can be written
in the conventional, "degree three" form.

\subsection{$W_3$ algebra}
Our first example of a quadratic algebra with an infinite number of generators is the $W_3$
algebra discovered by A.~Zamolodchikov in 1984 \cite{Zam1}. This algebra
contains the Virasoro algebra, generated by the stress-tensor $T(z)$, and has, in addition, a spin
3 current $W(z)$. In terms of the Operator Product Expansion (OPE) the defining relations for the
$W_3$ algebra have the form\footnote{The general form of the OPE's is $A(z_1)B(z_2)=(\mbox{pole terms})+(\mbox{non
singular terms})$. In what follows we will explicitly write only pole terms in all OPE's.}
\be\label{w3qu}
\begin{array}{rcl}T(z_1)T(z_2)&\sim &\displaystyle{\frac{c/2}{z_{12}^4}}+\displaystyle{\frac{2T}{z_{12}^2}}+
\displaystyle{\frac{T'}{z_{12}}}\ ,\quad T(z_1) W(z_2)\sim \frac{3 W}{z_{12}^2}+\displaystyle{\frac{W'}{z_{12}}}\ ,\\[1.2em]
W(z_1)W(z_2)&\sim&\displaystyle{\frac{c/3}{z_{12}^6}}+\displaystyle{\frac{2T}{z_{12}^4}}+
\displaystyle{\frac{T'}{z_{12}^3}}+
\displaystyle{\frac{a_1T'' +a(TT)}{z_{12}^2}}+
\displaystyle{\frac{a_2 T'''+\frac{a}{2} (TT)'}{z_{12}}}\ ,\end{array}\ee
where $z_{12}=z_1-z_2$ and
\be a=\frac{32}{22+5c}\ ,\ a_1=\frac{3c-6}{44+10c}\ ,\ a_2=\frac{2}{9}a_1\ .\ee
All currents in the r.h.s. of \p{w3qu} are evaluated at the point $z_2$ and all products of
currents are supposed to be normally ordered,
$(AB)(z)= \frac{1}{2\pi i} \oint_z d\zeta \frac{A(\zeta)B(z)}{\zeta -z}$.

\vskip .3cm
The main feature of the OPE's \p{w3qu} is the appearance of the quadratic combinations of the
currents $T$ in the OPE of the $W$ current with itself. These terms are absolutely necessary
to fulfill the Jacobi identity for the currents $T$ and $W$.

\vskip .3cm
The BRST charge for the $W_3$ algebra \p{w3qu} has been constructed by J.~Thierry-Mieg in
\cite{TM}. Explicitly the BRST charge reads
\be\label{charge1} {\tilde Q} = \frac{1}{2i \pi} \oint dz\, Q(z)\ ,\ee
where the BRST current $Q(z)$ has the following form\footnote{As usual, the BRST current is
defined up to full derivatives which disappear after the integration in \p{charge1}.}
\be\label{brst1}\begin{array}{rcl}Q&=&(c_TT)+(c_WW)-(b_Tc_T'\, c_T)-\displaystyle{\frac{125}{1566}}
(b_Tc_W'''\, c_W)-(c_T b_W c_W'')\\[1em] &&-\displaystyle{\frac{25}{522}}(b_T'\, c_W''\, c_W)+
2(c_T'\, b_Wc_W)-\displaystyle{\frac{8}{261}}(Tb_Tc_W'\, c_W)\ .\end{array}\ee

The ghosts---anti-ghosts currents $(c_T,c_W,b_T,b_W)$ obey the standard OPE's:
\be\label{ghosts1}b_T(z_1)c_T(z_2)\sim\frac{1}{z_{12}}\ ,\quad b_W(z_1)c_W(z_2)\sim\frac{1}{z_{12}}
\ee
(all other OPE's for the ghosts---anti-ghosts are regular).

\vskip .3cm
The BRST charge \p{charge1} squares to zero,
\be\label{nil}{\tilde Q}{}^2=0\ ,\ee
if the central charge in the $W_3$ algebra \p{w3qu} is equal to its critical value $c=100$.
With this central charge, the ghost modified current ${\tilde T}$,
\be\label{ghostT}{\tilde T}=T+T_{gh}=T-(b_T'\, c_T)-2(b_T c_T')-2(b_W'\, c_W)-3(b_W c_W')\ ,\ee
obeys the Virasoro algebra with the zero central charge
\be{\tilde T}(z_1){\tilde T}(z_2)\sim \frac{2{\tilde T}}{z_{12}^2}+\frac{{\tilde T}{}'}{z_{12}}\ .
\ee
One checks that with respect to ${\tilde T}$ the rest of the currents are primary, with the
conformal dimensions
\be\label{dim}[W]=3\, ,\, [c_T]=-1\ ,\ [b_T]=2\ ,\ [c_W]=-2\ ,\ [b_W]=3\ ,\ee
as it should be.

\vskip .3cm
The OPE's \p{w3qu}, \p{ghosts1} and the
BRST current \p{brst1} are invariant under the following automorphism transformations
\be\lb{auto}(T,c_T,b_T)\rightarrow (T,c_T,b_T)\ ,\quad (W,c_W,b_W)\rightarrow (-W,-c_W,-b_W)\ .\ee

\vskip .3cm
Observe that the last term in the expression \p{brst1} for the BRST charge is unconventional,
it has degree 4 in generators. Motivated by the discussions in the previous Section, we modify the
ghost---anti-ghost algebra and study the resulting consequences for the BRST charge.
The only restrictions we impose on the modified ghosts are: $(i)$ the relations remain quadratic;
$(ii)$ the compatibility with the conformal weights \p{dim}; $(iii)$ the invariance under the
automorphism \p{auto}. One checks that the most general non-linear ghosts---anti-ghosts algebra,
fulfilling these requirements, depends on two arbitrary parameters $(g_1, g_2)$. The corresponding
OPE's read:
\be\label{ghostq}\begin{array}{rcl}&&\tb_T(z_1)\tc_T(z_2)\sim\displaystyle{\frac{1}{z_{12}}}\ , \quad\tb_W(z_1)\tc_W(z_2)\sim\displaystyle{\frac{1}{z_{12}}}\ ,\\[1em]
&&\tc_T(z_1)\tb_W(z_2)\sim\displaystyle{\frac{g_1(\tb_T \tc_W)}{z_{12}^2}}+
\displaystyle{\frac{g_2\left(\tb_T\tc_W\right)'+g_1(\tb_T\tc_W')}{z_{12}}}\ ,\\[1em]
&&\tc_T(z_1)\tc_T(z_2)\sim\displaystyle{\frac{\left( g_1+g_2\right) (\tc_W'\,\tc_W)}{z_{12}}}\ ,
\quad\tb_W(z_1)\tb_W(z_2)\sim\displaystyle{\frac{\left( g_1-g_2\right) (\tb_T'\,\tb_T)}{z_{12}}}
\ .\end{array}\ee
The relations \p{ghostq}
define a quadratic algebra and the Jacobi identity is satisfied for arbitrary values of the
parameters $g_1$ and $g_2$.

\vskip .3cm
Now this is a matter of calculations to check that
the BRST current for the $W_3$ algebra
with the new ghosts system  $(\tc_{T,W},\tb_{T,W})$ exists for arbitrary values of the
parameters $g_1$ and $g_2$. The BRST current has the form
\be\label{brstq}\begin{array}{rcl}Q&=&(\tc_TT)+(\tc_WW)-(\tb_T\tc_T'\,\tc_T)-\left[ \displaystyle{\frac{125}{1566}}+\displaystyle{\frac{17}{12}}(g_1+g_2)\right]
(\tb_T\tc_W'''\,\tc_W) \\[1em]
&&-(\tc_T\tb_W\tc_W'')-\left[\displaystyle{\frac{25}{522}}+\frac{5}{4}(g_1+g_2)\right]
(\tb_T'\,\tc_W''\,\tc_W)+2(\tc_T'\,\tb_W\tc_W)\\[1em]
&&-\left[\displaystyle{\frac{8}{261}}+\displaystyle{\frac{1}{2}}(g_1+g_2)\right] (T\tb_T\tc_W'\,
\tc_W)-g_1(\tb_T'\,\tb_T\tc_T\tc_W'\,\tc_W)\ .\end{array}\ee
The corresponding BRST charge obeys the nilpotency condition \p{nil} if the
central charge in the algebra \p{w3qu} equals $c=100$ (independently of the values of the
parameters $g_1$ and $g_2$), as it should be.

\vskip .3cm

For general values of the parameters $g_1$ and $g_2$, the BRST current contains unconventional
terms -- the third line in \p{brstq}. The unconventional terms can be removed if we choose (and
this choice is unique)
\be\label{sol1}g_1=0\ ,\quad g_2=-\frac{16}{261}\ .\ee
Thus we see that the net effect of using the non-canonical ghost algebra is a freedom in the BRST
charge. This freedom could  be further fixed in a such way as to get the  conventional form of
BRST charge. Clearly, the standard, canonical, construction corresponds to the choice $g_1=g_2=0$.

\vskip .3cm
{}Finally, it is worth to note that for any values of the parameters $g_1$ and $g_2$ there is
a non-linear transformation sending the old, canonical, ghosts---anti-ghosts currents
(\ref{ghosts1}) to the modified ghosts---anti-ghosts currents (\ref{ghostq}).
It has the following form:
\be\lb{transf}b_T=\tb_T\ ,\ c_T=\tc_T-\frac{g_1+g_2}{2}(\tb_T\tc_W'\,\tc_W)\ ,\
b_W=\tb_W-\frac{g_1-g_2}{2}(\tb_T'\,\tb_T\tc_W)\ ,\ c_W=\tc_W\ .\ee
Although the transformation (\ref{transf}) is non-linear, it is clearly invertible.

Note that under this transformation (for $g_1=0$) the form of ghost stress-tensor
$$T_{gh}=-(\tb_T' \,\tc_T)-2(\tb_T \tc_T')-2(\tb_W'\,\tc_W)-3(\tb_W \tc_W')$$
does not change.

\subsection{$W_3^{(2)}$ algebra}
Another example of the quadratic algebra with an infinite number of generators is provided by the
so called $W_3^{(2)}$ algebra. In this subsection we demonstrate, without going into details, that
the corresponding BRST current can be also brought to the conventional form by a proper
modification of the ghost algebra.

\vskip .3cm
The $W_3^{(2)}$ algebra \cite{P}, \cite{B} is the bosonic analog of the well known $N=2$ super
Virasoro algebra. It contains four bosonic currents $T,U,G^+,G^-$ with conformal dimensions
$$[T]=2\ ,\ [U]=1\ ,\ [G^+]=[G^-]=3/2$$
which obey the following OPE's
\be\lb{w32}\hspace{-.2cm}\begin{array}{l}T(z_1)T(z_2)\sim\displaystyle{\frac{c(7-9c)}{2(1+c)}
\frac{1}{z_{12}^4}+\frac{2T}{z_{12}^2}+\frac{T'}{z_{12}}}\ ,\\[1em]
T(z_1)U(z_2)\sim\displaystyle{\frac{U}{z_{12}^2}+\frac{U'}{z_{12}}\ ,\
T(z_1)G^\pm(z_2)\sim\frac{\frac{3}{2}G^\pm}{z_{12}^2}+\frac{G^\pm{}'}{z_{12}}}\ ,\\[1em]
G^+(z_1)G^-(z_2)\sim\displaystyle{\left[\frac{2c-6c^2}{1+c}\right]\frac{1}{z_{12}^3}+
\left[\frac{2-6c}{1+c}\right] \frac{U}{z_{12}^2}+
\frac{2T-\frac{4}{1+c} (U\,U)+\frac{1-3c}{1+c}\,U'(z_2)}{z_{12}}}\ ,\\[1em]
U(z_1)G^\pm(z_2)\sim\displaystyle{\pm\frac{G^\pm}{z_{12}}}\ ,\
U(z_1)U(z_2) \sim \frac{c}{z_{12}^2}\ .\end{array}\ee
The BRST charge for this non-linear algebra has been constructed in \cite{ssn}, \cite{brstw32}.
It obeys the nilpotency condition \p{nil} for the critical value of the central charge $c=-2$,
which corresponds to the Virasoro subalgebra central charge\footnote{The standard normalization
for the Virasoro algebra reads $T(z_1)T(z_2) \sim \frac{c_{Vir}/2}{z_{12}^4}+\ldots$.}
$c_{Vir}=50$.

\vskip .3cm
In a full analogy with the $W_3$ algebra, considered in the previous subsection, one
checks that there is a conventional BRST current
\be\label{brst32}\begin{array}{l}Q=(c_TT)+(\tilde{c}_U)U+(c^+G^+)+(c^-G^-)+(\tilde{c}_U b^+c^+)-
(\tilde{c}_Ub^-c^-)+\frac{1}{2}(c_T'\, b_U\tilde{c}_U)\\[.5em]
\hspace{.5cm}\displaystyle{+\frac{1}{2}(c_T b_U'\,\tilde{c}_U)-\frac{1}{2}(c_Tb_U\tilde{c}_U')
+\frac{3}{4}(c_T'\, b^+c^+)+\frac{1}{4}(c_Tb^+{}'\, c^+)-\frac{3}{4}(c_Tb^+c^+{}')}\\[.7em]
\hspace{.5cm}\displaystyle{+\frac{3}{4}(c_T'\, b^-c^-)+\frac{1}{4}(c_Tb^-{}'\, c^-)-\frac{3}{4}
(c_Tb^-c^-{}')+4(b_Uc^+c^-{}')+3(b_U'c^+c^-)}\\[.6em]
\hspace{.5cm}\displaystyle{+2(b_Uc^+{}'\,c^-)-(\tilde{b}_Tc_T'\, c_T)-
2(\tilde{b}_Tc^+c^-)}\ \end{array}\ee
if the ghost---anti-ghost currents form a quadratically non-linear algebra
\be\label{gagh2}\begin{array}{l}\displaystyle{\tilde{b}_T(z_1)c_T(z_2)\sim\frac{1}{z_{12}}\ ,\
b_U(z_1)\tilde{c}_U(z_2)\sim\frac{1}{z_{12}}\ ,\
b^\pm (z_1)c^\pm(z_2)\sim\frac{1}{z_{12}}}\ ,\\[.8em]
\displaystyle{\tilde{b}_T(z_1)\tilde{c}_U(z_2)\sim -2\frac{(c_Tb_U)}{z_{12}^2}-
2\frac{2(c_Tb_U')+(c_T'\, b_U)}{z_{12}}}\ ,\\[.8em]
\displaystyle{\tilde{b}_T(z_1)\tilde{b}_T(z_2)\sim -4\frac{(b_U'\, b_U)}{z_{12}}\ ,\
\tilde{c}_U(z_1)\tilde{c}_U(z_2)\sim -8\frac{(c^+c^-)}{z_{12}}}\ , \\[.8em]
\displaystyle{\tilde{c}_U(z_1)b^+(z_2)\sim 4\frac{(b_Uc^-)}{z_{12}}\ , \
\tilde{c}_U(z_1)b^-(z_2)\sim -4\frac{(b_U c^+)}{z_{12}}}\ . \end{array}\ee

\noindent
As for the $W_3$ algebra, one can relate the ghost---anti-ghost currents $(c_T,\tilde{b}_T,
\tilde{c}_U,b_U,c^\pm,b^\pm)$ obeying the OPE's \p{gagh2} to the canonical ones $(c_T,
b_T, c_U, b_U, c^\pm, b^\pm)$, obeying the standard OPE's
\be b_T(z_1)c_T(z_2)\sim\frac{1}{z_{12}}\ ,\ b_U(z_1)c_U(z_2)\sim\frac{1}{z_{12}}\ ,\
b^\pm (z_1)c^\pm(z_2)\sim\frac{1}{z_{12}}\ ,\ee
by a non-linear invertible transformation
\be\label{modghost}\tilde{b}_T=b_T-2(c_Tb_U'\, b_U)\ ,\ \tilde{c}_U=c_U-4(b_Uc^+c^-)\ \ee
(only the currents $b_T$ and $c_U$ get transformed).

\setcounter{equation}0
\section{Conclusion}

In this paper we have studied BRST operators for quadratic algebras. We have shown that the form
of the BRST operator for the $W_3$ and $W_3^{(2)}$ algebras can be considerably simplified if one
allows for a non-canonical quadratic ghost---anti-ghost sector instead of the canonical one.
The possibility of having non-canonical ghosts---anti-ghosts is motivated by the BRST theory for
Lie super-algebras (discussed in the Introduction) as well as by the Woronowicz theory \cite{Wor}
of differential calculi on quantum groups where the algebra of differential forms is deformed
according to the algebra of non-commutative vector fields (the BRST operator in this case was
constructed in \cite{IsOg2}, \cite{IsOgGo}). The vector fields in the differential calculus
on quantum groups obey the reflection equation algebra (in dimension 2 this algebra is the 
$q$-Minkowski space, \cite{OSWZ}) which is generated by 
entries of a matrix $L$ with the relations
\be L_1\hat{R}_{12}L_1\hat{R}_{12}=\hat{R}_{12}L_1\hat{R}_{12}L_1
\ .\label{rea}\ee
Here $\hat{R}$ is an arbitrary Yang--Baxter operator. If $\hat{R}^2$ is not proportional
to one, the algebra (\ref{rea}) acquires linear terms upon the substitution 
$L\longrightarrow L+\alpha\, {\mathrm{Id}}$, where 
$\alpha\neq 0$ is an arbitrary constant. The reflection equation algebra (with the shifted generators) 
is an example of a quantum Lie algebra; the ordering relations for copies of generators 
have the following form
\be L_1^{(A)}L_{\overline{2}}^{(B)}=\hat{R}_{12}L_1^{(B)}L_{\overline{2}}^{(A)}\hat{R}_{12}^{-1}
+\alpha \Bigl( \hat{R}_{12}^2 L_1^{(A)}\hat{R}_{12}^{-2}-L_1^{(A)}\Bigr)\qquad\ {\mathrm{for}}
\quad A<B\ ,\label{reac}\ee
where $L_{\overline{2}}:=\hat{R}_{12}L_1\hat{R}_{12}^{-1}$; the upper index in $L^{(A)}$ labels 
the copy. The set (\ref{int1a})-(\ref{int3a}) of the QLA identities for $\sigma^{ij}_{kl}$ and 
$C^i_{jk}$ is equivalent to the equality of two different reorderings of 
$\chi_{1\rangle}^{(A)}\chi_{2\rangle}^{(B)}\chi_{3\rangle}^{(C)}$
with $A<B<C$ (see \cite{IsOg2} for details). To verify the QLA identities for the relations (\ref{reac}), 
it is convenient to reorder the combination 
$L_1^{(A)}L_{\overline{2}}^{(B)}L_{\overline{3}}^{(C)}$, where
$L_{\overline{3}}:=\hat{R}_{23}L_{\overline{2}}\hat{R}_{23}^{-1}$. The BRST 
operator for the reflection equation algebra was constructed in \cite{IsOg5}, \cite{BIO}.
 
\vskip .3cm
We note that the phenomenon of the non-linear realization of the non-canonical ghosts via the
canonical ones, observed in Section 3 for the $W$-algebras, is not universal and can not,
in general, be realized for the quadratic algebras from Section 2, even if the number of
generators is finite (cf. \cite{Og} where it is shown that the standard Drinfeld-Jimbo deformation 
of the Heisenberg relations leads to the algebra, which is isomorphic, up to a certain completion,
to the non-deformed one).

\vskip .3cm

Strictly speaking, the $W$-algebras from Section 3 are not defined in the form appropriate
for the quantum Lie algebras; we don't know the braid matrix formulation for these algebras.

\vskip .3cm
An intriguing question is how to write a non-linear ghost sector for a general non-linear algebra.
For the $W$-algebras in particular, the appropriate non-linear ghost sector has to follow from the
basic OPE's of the algebra we started from. An answer to this question could help to solve long
standing problems of construction of the BRST charges for some infinitely generated
super-algebras. In this respect the most interesting case is the structure of the BRST charge for
the $N=2$ super-symmetric $W_3$ algebra \cite{N2}, \cite{N3}. The main problem here is the infinite
Ansatz for the BRST charge (within the standard formulation) due to the
presence of bosonic ghosts---anti-ghosts combinations with zero conformal weight and zero ghost
number. We hope that the present approach will help in solving this problem, which probably
will open a way to construct $N=4$ super $W_3$ algebra along the lines presented in \cite{N4}.

\section*{Acknowledgements}

We are grateful to I. Buchbinder, R. Coquereaux, P. Lavrov, Ya. Pugai and A. Zamolodchikov
for valuable discussions.

\vskip .2cm
A part of this work was done while two of us (A.P.I. and S.O.K.) were
visiting Marseille University and Centre de Physique Th\'eorique (Luminy, Marseille). We thank
Marseille University and Centre de Physique Th\'eorique for the financial support. A.P.I. is also
grateful to  Centre International de Rencontres Math\'ematiques (Luminy, Marseille) for the kind
hospitality and support. The work of A.P.I. was also partially supported by the grant
RFBR-08-01-00392-a. The work of S.O.K. was partially supported by  INTAS under contract
05-7928 and by grants RFBR-06-02-16684, 06-01-00627-a,  DFG~436 Rus~113/669/03. The work of
O.V.O. was supported by the ANR project GIMP No.ANR-05-BLAN-0029-01.

\end{document}